\documentclass{article}\usepackage[]{graphicx}\usepackage[dvipsnames]{xcolor}

\makeatletter
\def\maxwidth{ %
  \ifdim\Gin@nat@width>\linewidth
    \linewidth
  \else
    \Gin@nat@width
  \fi
}
\makeatother

\definecolor{fgcolor}{rgb}{0.345, 0.345, 0.345}

\usepackage{authblk}
\usepackage{framed}
\makeatletter
 {\par\unskip\endMakeFramed%
 \at@end@of@kframe}
\makeatother

\definecolor{shadecolor}{rgb}{.97, .97, .97}
\definecolor{messagecolor}{rgb}{0, 0, 0}
\definecolor{warningcolor}{rgb}{1, 0, 1}
\definecolor{errorcolor}{rgb}{1, 0, 0}

\usepackage{alltt}

\title{Non-linear dependence and Granger causality:\\ A vine copula approach}
\author[1,2]{Roberto Fuentes-Martínez}
\author[1]{Irene Crimaldi}
\author[1]{Armando Rungi}
\date{}
\affil[1]{\small IMT School for Advanced Studies Lucca}
\affil[2]{\small Universidad de Alicante}
\usepackage[english]{babel}
\usepackage[utf8]{inputenc}
\usepackage[a4paper, width=150mm,top=25mm,bottom=25mm]{geometry}
\usepackage{mathpazo}
\usepackage{setspace}

\usepackage{graphicx}
\usepackage{appendix}
\usepackage{epigraph}
\usepackage{longtable}
\usepackage{array}
\usepackage{caption}
\usepackage{subcaption}
\usepackage{hyphenat}
\usepackage{caption}
\usepackage{multirow}
\usepackage{longtable}
\usepackage{makecell}
\usepackage{bbding}
\usepackage{booktabs}
\usepackage{dcolumn}
\usepackage{algorithm}
\usepackage{bm}
\usepackage{amsmath}
\usepackage{mathrsfs}
\usepackage{tikz}
\usepackage{tabularx}
\usetikzlibrary{calc}
\usepackage{tabularray}
\usepackage{multirow,array}
\usepackage{comment}

\usepackage{amsthm}
\usepackage{amssymb}
\usepackage{bm}
\usepackage[bbgreekl]{mathbbol}
\usepackage{xspace}
\usepackage{afterpage}

\usepackage{booktabs} 
\usepackage{float}
\usepackage{multirow}
\usepackage{mathtools}
\usepackage{natbib}
\usepackage[flushleft]{threeparttable}
\newtheorem{definition}{Definition}
\newtheorem{theorem}{Theorem}
\usepackage[T1]{fontenc}
\usepackage{lmodern}
\IfFileExists{upquote.sty}{\usepackage{upquote}}{}
\begin{document}

\newpage 
\newpage
\setcounter{page}{1}
\date{\today}
\maketitle
\begin{abstract}
Inspired by \cite{jang2022}, we propose a Granger causality-in-the-mean test for bivariate $k-$Markov stationary processes based on a recently introduced class of non-linear models, i.e., vine copula models. By means of a simulation study, we show that the proposed test improves on the statistical properties of the original test in \cite{jang2022}, and also of other previous methods,  constituting an excellent tool for testing  Granger causality in the presence of non-linear dependence structures. Finally, we apply our test to study the pairwise relationships between energy consumption, GDP and investment in the U.S. and, notably, we find that Granger-causality runs two ways between GDP and energy consumption.  
\end{abstract}

\section{Introduction}

The notion of Granger causality was seminally introduced by \cite{granger1969}, and then usefully applied in several scientific fields, including economics, finance, neuroscience, genomics, and climate science \citep{shojaie2022}, where the characterization of the dependence relations between time series is a relevant issue. It is based on the comparison of the quality of the prediction when forecasting future values of one series with or without the information about the past values of the other time series. In other words, a time series has a Granger causal influence on another time series if its past values facilitate the prediction of the other time series. Therefore, briefly, testing Granger causality means checking if there is an improvement in prediction if we leverage not only the past values of the considered series but also the past values of the other series.\\

In this paper, we propose a novel Granger causality-in-the-mean test for bivariate $k-$Mar\-kov stationary processes, which is based on vine copula models. In general, copula functions are the mathematical tool that describes any type of dependence structure and so they are able to capture non-linear dependencies. In particular, vine copulas are highly flexible models that allow for the representation of a wide array of different joint distributions \citep{czado2022} and computationally tractable model-selection and estimation procedures. Formally, exploiting the concept of \textit{pair copula construction}  \citep{joe1996}, a multivariate copula is constructed using bivariate copulas as building blocks, overcoming the limited and restrictive selection of multivariate copulas. As shown in recent literature \citep{brechmannczado2015,beare2015,smith2015,nagler2022}, vine copulas are non-linear models that can represent multivariate time series by capturing both their cross-sectional and serial dependence within the same model, giving place to a myriad of applications in the context of time series modelling, one of them being the study of Granger causality. \\

\indent In literature, there are already several tests differing from each other in the models and techniques used for measuring and estimating the correspondent measure of Granger causality. Gran\-ger's original formulation and most of the existing tests are based on linear models such as Vector Autoregressive (VAR) models. Yet, the application of these tools to non-linear systems may not be appropriate \citep{ancona2004}. Therefore, a few non-linear Granger causality tests have been proposed \citep{hiemstra1994,diks2006,diks2016, song2018, kim2020}. In economics, the \cite{hiemstra1994} test (HJ test hereafter) has been the most popular approach for testing non-linear Granger causality \citep{bai2017}; however, it has been shown through simulation studies that this test has a tendency to over-reject the null hypothesis. \cite{diks2006} proposed a modified version of the HJ test that overcomes the rejection rate issues by changing the test statistics and providing some guidelines for the optimal bandwidth selection based on the sample size. Afterwards, \cite{diks2016} extended the setting by \cite{diks2006} from a bivariate to a multivariate one. \cite{song2018} introduced a model-free measure of non-linear Granger causality that takes the form of a log-difference between restricted and unrestricted mean squared errors, building a test by means of a non-parametric estimate of this measure using kernel estimators of mean squared errors. \cite{hiemstra1994}, \cite{diks2006}, \cite{diks2016} and \cite{song2018} introduce non-parametric methods. More recently, exploiting the measure of \cite{song2018}, \cite{jang2022} proposed a semi-parametric Granger causality test based on vine copulas, which can capture non-linear dependencies with a good statistical performance in terms of size and power if compared to the aforementioned tests. In line with this recent literature, our aim is to propose a test that extends and improves on the family of semi-parametric non-linear Granger causality tests, which statistical properties are not heavily dependent on the choices of bandwidth or smoothing parameters, as it is instead the case for most of the existing non-parametric non-linear Granger causality tests.\\

\indent In this contribution, we start from the test procedure proposed in \cite{jang2022} and incorporate some relevant methodological differences. Thus, we build our non-linear Granger causality-in-the-mean test for bivariate $k-$Markov stationary processes, which we call M-vine Granger causality test. By means of a simulation study based on the setting from \cite{song2018}, we find that the M-vine test has a higher power than the original version from \cite{jang2022} in all scenarios considered, while still controlling its size under or considerably close to the predefined significance level. Moreover, for linear specifications and large samples, our test displays a power that resembles the one from a traditional Granger-causality test based on linear models, even if the latter is based on the real data generating process. Yet, for non-linear specifications, the M-vine test has a substantially higher power than the linear Granger-causality test. Hence, we show that our test improves on the statistical properties of the original one in \cite{jang2022}, and also of other previous methods, and makes it an excellent tool for testing  Granger causality in the presence of non-linear dependence structures.\\

\indent Finally, we apply the M-vine Granger causality test to study the pairwise causal relationships between GDP, energy consumption and investment in the U.S. The nexus between energy consumption and economic growth has been studied in various works   \citep{kraft1978, chen2007, payne2009, ozturk2010}. Nonetheless, results appear to be diverse and often conflicting, even when they refer to the same country: for instance, in Appendix~\ref{prevlit}, Table~\ref{table-literature}, we provide a synthetic recap of the results from different studies that specifically investigated Granger causality between energy consumption and economic growth with a focus on the United States. Here, we provide our contribution by means of the M-vine test, and we compare results with the S-vine test and a more classical linear test based on VARs. At $5\%$ significance level, we find a Granger causality relationship that runs from GDP to energy consumption, which is in line with previous literature \citep{kraft1978, fallahi2011, aslan2014}). Yet, at the same time, we also find that there is an albeit less significant Granger causality running from energy consuption to economic growth that is still in line with other pieces of economic literature \citep{stern1993, stern2000, bowden2009, fallahi2011}. Crucially, only our M-vine test is able to catch the two-way Granger causality between energy consumption and GDP.
\\

\indent The rest of the paper is organized as follows. Section~\ref{method} introduces the methodology and the setting of the M-vine test, while section~\ref{simst} studies the statistical properties of the proposed test by means of a simulation study. In section~\ref{empapp}, we present a macroeconomic application to analyse the pairwise relationships between energy consumption, GDP and investment in the U.S. Conclusions and final remarks are offered in Section~\ref{conc}.

\section{Methodology}\label{method}

Measures of Granger causality are classified into three types, according to the criterion chosen to measure their prediction quality: namely, Granger causality in the mean, in the quantiles or in distribution. Here, we focus only on the first, which is the most widely used in literature, while we refer to \cite{lee2014} and \cite{candelon2016} for the others. Nonetheless, we believe it is important to remark that the methodology that we present can be naturally extended to Granger causality in quantiles \citep{taamouti2021,jang2023}.
\\

\indent Let $(X_t, Y_t)_{t\in\mathbb{Z}}$ be a strictly stationary bivariate stochastic process, that is 
$$
[(X_t, Y_t),\dots, (X_{t+h}, Y_{t+h})] \stackrel{d}=
[(X_s, Y_s),\dots, (X_{s+h}, Y_{s+h})]\qquad\forall s,\,t\in\mathbb{Z},\, h\in\mathbb{N}
$$ 
where the symbol $\stackrel{d}=$ means equality in distribution. 
Suppose also that $(X_t, Y_t)_{t\in\mathbb{Z}}$ is a Markov process of order $k\in \mathbb{N}\setminus\{0\}$ (more briefly, a $k$-Markov process), that is 
$$
[X_{t},Y_{t}] \,|\, [(X_s,Y_s):\,s\leq t-1]\stackrel{d}= [X_{t},Y_t] \,|\, [(X_{t-1},Y_{t-1}),\dots, (X_{t-k},Y_{t-k})]
\qquad\forall t\,.
$$ 
Set $X=(X_t)_{t\in\mathbb{Z}}$ and $Y=(Y_t)_{t\in\mathbb{Z}}$.

\begin{definition}{\textbf{(Granger causality in the mean)}}
Assuming $X_t$ and $Y_t$ to be square integrable, we say that $Y$ causes $X$ in the sense of Granger, if 
\begin{equation*}
\sigma^2(X_t|X_s,Y_s:\, s=t-1,\dots,t-k) < \sigma^2(X_t|X_s:\,s=t-1,\dots, t-k)
\end{equation*}
where 
\begin{equation*}
\begin{split}
&\sigma^2(X_t|X_s:\,s=t-1,\dots, t-k)=\mathbb{E}\left[(X_t-\mathbb{E}[X_t|X_s:\,s=t-1,\dots, t-k])^2\right]
\quad\mbox{and}\\
&\sigma^2(X_t|X_s,Y_s:\, s=t-1,\dots,t-k)=\mathbb{E}\left[(X_t-\mathbb{E}[X_t|X_s,Y_s:\, s=t-1,\dots,t-k])^2\right]
\end{split}
\end{equation*} 
are the mean squared errors of the optimal prediction of $X_t$ 
given the past observations $[X_{t-1},\dots, X_{t-k}]$ of $X$ and given the past observations  $[X_{t-1},Y_{t-1},\dots,X_{t-k},Y_{t-k}]$ of both $X$ and $Y$, respectively.
\end{definition}
\indent A measure for Granger causality in the mean has been proposed in \cite{song2018} as  
\begin{equation*}
GC^{mean}(Y\rightarrow X)=\log 
\left[
\frac
{\sigma^2(X_t|X_s:\,s=t-1,\dots, t-k)}
{\sigma^2(X_t|X_s,Y_s:\, s=t-1,\dots,t-k)}
\right].
\end{equation*}
Note that $GC^{mean}(Y\rightarrow X)$ is a non-negative quantity, which is equal to zero when there is no Granger causality in the mean (i.e.~when the two above mean squared errors are equal).  
Moreover, it is not restricted to a given model for the distributions of the involved random variables. 
This measure can be used to construct a statistical test in order to check the presence or absence of Granger causality in the mean  
 for two $k$-Markov stationary time series. \\
 
\indent \cite{jang2022} proposed a semi-parametric 
 Granger causality test based on the above measure $GC^{mean}(Y\rightarrow X)$, estimated using 
  a {\em copula approach} (see Appendix~\ref{copapp}). Copula functions are the mathematical tool that describes any type of dependence structure. Therefore,  this test is able to capture non-linear dependence and, moreover, it  exhibits good statistical properties when compared to other existing non-linear tests (see \cite{jang2022}). Hence, it is our starting point.
  We here propose a test for Granger causality in the mean 
  based on the M-vine copula structure presented in \cite{beare2015} for a 
  bivariate $k-$Markov stationary stochastic process. The presented test improves both on the size and the power when compared to the 
  one in \cite{jang2022}, and it is more suitable for empirical applications 
  since it takes into account possible non-linear dependencies that are commonly encountered when working with real data. Please, refer to Appendix~\ref{copapp} for some recalls on the notion of copulas and, in particular, on vine copulas.

  \subsection{M-vine Granger causality test}\label{mvtest}

Let $(X,Y)=(X_t,Y_t)_t$ be a bivariate $k-$Markov square-integrable stationary stochastic process and let $\{(x_t,y_t)\;:\;t=1,\dots,T\}$ be a sample of it.
  To simplify notation, from now on, 
  we will assume $k=1$, but everything can be naturally extended to any $k\in\mathbb{N}\setminus\{0\}$.
  \\
  
  \indent We can split the proposed test, which we call the M-vine Granger causality test, into two parts, say part A and part B: the first one regards the evaluation of an estimator  of the Granger causality (in the mean) measure, which is used as the test statistics, and the second one  concerns the computation of the $p$-value of the test.
  \\
  
\noindent We start by describing part 
A (computation of the value of the test statistic). We use the sample version of the above-recalled measure $GC^{mean}(Y\rightarrow X)$, that is 
 \begin{equation*}
 \begin{split}
\widehat{GC}^{mean}_{Y\rightarrow X}&=
\log \left[
\frac{(T-T_0+1)^{-1}\sum^T_{t=T_0}
\left(  x_t - \widehat{\mathbb{E}}[X_t|X_{t-1}=x_{t-1}] \right)^2}
{(T-T_0+1)^{-1}\sum^T_{t=T_0}
\left(  x_t - \widehat{\mathbb{E}}[X_t|X_{t-1}=x_{t-1},Y_{t-1}=y_{t-1}] \right)^2}
\right]\\
&=
\log \left[
\frac{\sum^T_{t=T_0}
\left(  x_t - \widehat{\mathbb{E}}[X_t|X_{t-1}=x_{t-1}] \right)^2}
{\sum^T_{t=T_0}
\left(  x_t - \widehat{\mathbb{E}}[X_t|X_{t-1}=x_{t-1},Y_{t-1}=y_{t-1}] \right)^2}
\right]\,,
\end{split}
\end{equation*}
where the quantities $\widehat{\mathbb{E}}[X_t|X_{t-1}=x_{t-1}]$ and  
$\widehat{\mathbb{E}}[X_t|X_{t-1}=x_{t-1}, Y_{t-1}=y_{t-1}]$ are computed by fitting an M-vine copula model to the data (see Appendix~\ref{sec-M-vine}) and using it to generate i.i.d. observations of $X_t$ given $X_{t-1}=x_{t-1}$ and of $X_t$ given $X_{t-1}=x_{t-1}$ and $Y_{t-1}=y_{t-1}$ and using the corresponding empirical means to estimate the needed conditional expectations.  
%
%
More precisely, we proceed as follows:
\begin{itemize}
\item[Step A1)] we fit an M-vine copula model and estimate its respective parameters for the observations from the series $X$, that is for  $M_x=(x_t)_{t=1\dots,T}$ we adopt the selection and estimation procedure introduced and studied in \cite{nagler2022}, where, among other things, its consistency is proven under suitable assumptions; see Section~\ref{sec-estimation-details} for details);\\ 
\item[Step A2)] using the model obtained from Step A1), for each $t=T_0,\dots, T$,  we 
generate $N$ i.i.d. predictions $\{\tilde{x}^{M_x}_{i,t}\}_{i=1}^N$ of $X_t$ given $X_{t-1}=x_{t-1}$, so that we can estimate the conditional mean of $X_t$ given $X_{t-1}=x_{t-1}$ by the empirical mean of the sample $\{\tilde{x}^{M_x}_{i,t}\}_{i=1}^N$, that is we set 
\begin{equation*}
\widehat{\mathbb{E}}[X_t|X_{t-1}=x_{t-1}]= \frac{1}{N}\sum_{i=1}^N \tilde{x}^{M_x}_{i,t}\qquad t=T_0,\dots,T;
\end{equation*}
\item[Step A3)] repeat Step A1) and Step A2) for the observations from the bivariate series $(X,Y)$, 
that is for  $M_{xy}=(x_t,y_t)_{t=1\dots,T}$, in order to obtain 
\begin{equation*}
\widehat{\mathbb{E}}[X_t|X_{t-1}=x_{t-1},Y_{t-1}=y_{t-1}]= 
\frac{1}{N}\sum_{i=1}^N \tilde{x}^{M_{xy}}_{i,t} \qquad t=T_0,\dots,T.
\end{equation*}
\end{itemize}
Introducing the sample estimates of both conditional expectations in $\widehat{GC}^{mean}_{Y\rightarrow X}$, we obtain
\begin{equation*}
\widehat{GC}^{mean}_{Y\rightarrow X}=\log \left[
\frac
{\sum^T_{t=T_0}\left(  x_t -  \frac{1}{N}\sum_{i=1}^N \tilde{x}^{M_x}_{i,t} \right)^2}
{\sum^T_{t=T_0}\left(  x_t -  \frac{1}{N}\sum_{i=1}^N \tilde{x}^{M_{xy}}_{i,t} \right)^2}
\right].
\end{equation*}
Note that, by the strong law of large numbers, for $N\to +\infty$, the empirical means $\frac{1}{N}\sum_{i=1}^N \tilde{x}^{M_x}_{i,t}$ and $\frac{1}{N}\sum_{i=1}^N \tilde{x}^{M_{xy}}_{i,t}$ are strongly consistent estimators of the conditional mean of $X_t$ given $X_{t-1}=x_{t-1}$ and given $X_{t-1}=x_{t-1},\, Y_{t-1}=y_{t-1}$ respectively, with respect to the model obtained from Step A1. Therefore, the computed quantity  $\widehat{GC}^{mean}_{Y\rightarrow X}$ results to be a strongly consistent estimator (for $N\to +\infty$) of the Granger causality measure with respect to the model obtained from Step A1 applied to $M_x$ and $M_{xy}$).   
\\

Roughly speaking, $\widehat{GC}^{mean}_{Y\rightarrow X}$, measures the log-difference between the prediction error related to the model with an M-vine copula structure fitted only on the information set from $X$,  
and the prediction error computed with the M-vine copula model fitted to the entire sample 
form both $X$ and $Y$, that is  using the information set of both series. 
In line with the definition of Granger causality in the mean, if this estimated quantity is significantly 
higher than zero, we reject the null hypothesis of no Granger causality (in the mean) from $Y$ to $X$: indeed, under the null hypothesis, the measure $GC^{mean}(Y\rightarrow X)$ is zero, and so, the higher is the value of the 
statistics $\widehat{GC}^{mean}_{Y\rightarrow X}$, the more statistically significant is the evidence of the presence of 
Granger causality (in the mean) running from $Y$ to $X$. Regarding $T_0\geq k+1$, 
we can choose it by taking into account the goodness of fit of the models to the data. 
\\

\indent In order to test if $\widehat{GC}^{mean}_{Y\rightarrow X}$ is statistically greater than zero, we rely, as in \cite{jang2022}, on a method, which takes inspiration from \cite{paparoditis2000}. 
This method relies on the M-vine copula model fitted on the entire sample $M_{xy}=\{(x_t,y_t):t=1,\dots,T\}$ in order to 
generate independent samples under the null hypothesis of no Granger causality (in the mean) and use them for computing the $p-$value for the test. 
 More precisely, the part B (computation of the $p$-value) of the proposed procedure works as follows:
\begin{itemize}  
\item[Step B1)]  from the first tree of the obtained M-vine copula structure, extract, for each $t=1,\dots, T$, both the copula between
 $X_t$ and $X_{t+1}$, say $c_{X_t,X_{t+1}}$, related to the conditional distribution of $X_{t+1}$ given $X_t$, 
 and the copula between $X_t$ and $Y_t$, say $c_{X_t,Y_t}$, related to the conditional distribution of $Y_t$ given $X_t$;
\item[Step B2)] using the estimated marginal distribution $\widehat{F}_X$, generate $x_t^0$, and, conditional on this value
 draw $x_{t+1}^0$ from $c_{X_t,X_{t+1}}$; 
\item[Step B3)] using also the estimated marginal distribution $\widehat{F}_Y$,  
  conditional on $x_t^0$, draw $y_t^0$ from $c_{X_t,Y_t}$; 
\item[Step B4)] using this generated sample $\{(x_t^0,y_t^0):t=1,\dots,T\}$, 
compute the quantity $\widehat{GC}^{0\;mean}_{Y\rightarrow X}$, that gives a simulated value of the test statistic under the null hypothesis; 
  
\item[Step B5)] repeat the above steps $B$ times, so that we get $B$ independent simulated values under the null hypothesis:   $\widehat{GC}^{0\;mean}_{j\;Y\rightarrow X}$ for $j=1,\dots,B$;
\item[Step B6)] compute the $p$-value for the test by the empirical mean
\begin{equation*}
p=\frac{1}{B}\sum_{j=1}^B \mathbb{1}(\widehat{GC}^{0\;mean}_{j\;Y\rightarrow X} \geq \widehat{GC}^{mean}_{Y\rightarrow X}).
\end{equation*}
\end{itemize}
Therefore, we reject the null hypothesis of no Granger causality (in the mean) when $p<\alpha$, where $\alpha$ is a given significance level.

\subsection{Differences with the S-vine Granger causality test by \cite{jang2022} }
\label{sec-diff}
There are two relevant differences between the test by \cite{jang2022} and the one we propose. Firstly, in the above-described Step 1A), we choose the class of the M-vine copulas to fit the data and to compute the value of the test statistic. The same model is then used in the procedure (part B) to compute the $p$-value for the test.  On the contrary, \cite{jang2022} use a more general class of copula models (that includes the M-vine copulas), i.e. the  class of S-vine copulas, for fitting the data and computing the test statistic, but then they restrict to the M-vine copulas for the computation of the $p$-value. The justification of this fact given in \cite{jang2022} simply relies on the implementation of this last computation: indeed, to draw the samples in the above Steps B2) and B3), both $c_{X_{t},X_{t+1}}$ and $c_{X_t,Y_t}$ are needed, and these two copulas are guaranteed to be present in the first tree of a given M-vine structure, whilst these two copulas might not appear in an S-vine structure (see Section~\ref{sec-M-vine} for more details). Consequently, in \cite{jang2022}, the conditional distributions used for generating a sample of the Granger causality  measure under the null hypothesis might be different from the ones in the model fitted to the data to estimate the Granger causality measure, i.e. to compute the value of the test statistic. In our procedure, the two parts, A (computation of the test statistic) and B (computation of the $p$-value), of the Granger causality test are aligned: they rely on the same class of vine copulas, i.e.~the M-vine copulas. 
\\

\indent Secondly, in the above Step 1A), we fit the M-vine copula model on the entire sample, i.e., on all the $T$ observations, while \cite{jang2022} split the entire sample into a training set, say observations from $t=1$ to $t=T^*$  (usually $T^*=T/2$), and a testing set, say observations from $t=T^*+1$ to $t=T$. Therefore, the fit of the $S$-vine copula model is performed only on the training set, while the testing set is used to compute the Granger causality statistics. The latter choice implies that all the predictions 
$\{\tilde{x}^{M_x}_{i,t}\}_{i=1}^N$ and $\tilde{x}^{M_{xy}}_{i,t}$ for $t=T^*+1,\dots, t$, used to compute the Granger causality statistics are based on a model fitted only on the observations from $t=1,\dots, T^*$. In real applications, when the time series are not perfectly stationary or the sample is small, this way of proceeding could generate poor estimates of the Granger causality measure. This is especially true in the case of non-linearities in the data generating process. In our case, instead, predictions are based on a model fitted to the entire sample, and this potentially makes our test better at detecting Granger causality, especially when the dependence structure is not linear. Note that fitting the model on more observations does not mean to add more model parameters, hence the risk of ``overfitting''  the data is not present.\\ 

\indent In the next subsection, we show that our variant has a better performance in terms of size and power.
 
\section{Simulation Study}\label{simst}

In this section, we analyse the statistical performance of our M-vine Granger causality test 
(briefly, M-vine test)  
and compare it, in terms of size and power,  
with the traditional linear Granger-causality test based on VAR models 
\footnote{The results of the linear test are obtained using the function \emph{grangertest} from the R package \emph{lmtest}. For the formulation of the linear test, we refer to Appendix~\ref{lintest}}, the S-vine Granger causality test (briefly, S-vine test) by \cite{jang2022}, and with a non-linear extension of the traditional Granger causality test using feed-forward neural networks from \cite{Hmamouche2020} (briefly, NlinTS)
\footnote{For the simulation study, we decided to compare the M-vine test only with methods for which the code is available, so that we could run the simulations ourselves and ensure the validity of the results.}. We recall that, as shown in \cite{jang2022}, the S-vine test has a good statistical performance in terms of size and power 
when compared to other non-linear Granger causality tests \citep{diks2006, song2018, kim2020}. 
Specifically, through a simulation study, they show that S-vine test, ST test \citep{song2018}  and KLH test \citep{kim2020}  are the ones whose size is closer to the predefined significance level, whereas the size of the DP test \citep{diks2006} is considerably smaller than the significance level and this suggests a tendency of this test to under-reject. Indeed, 
in terms of statistical power, \cite{jang2022} evidences that the DP test is the one that performs 
the worst. On the contrary, the S-vine test shows, in large samples ($T\geq 200$ for Markov processes of order $k=1$), a good power for all the dependence structure, while the KLH test has an excellent power only when working with linear or quasi-linear models and the ST test displays a good power exclusively when the dependence structure is non-linear. In the following, we will show in particular that our M-vine test outperforms the S-vine test in all the considered scenarios, exhibiting a very good power already starting from $T=100$ in the case of Markov order $k=1$.  \\

\indent We performed a simulation study based on the assessment models from \cite{jang2022}. We used their three size assessment models  (S1, S2 and S3) and their three power assessment models (P1, P2 and P3). 
Furthermore, we enhanced the exercise by including other data generating processes: two size assessment models (S4 and S5) and one power assessment  model (P5) taken from 
\cite{song2018} and  an additional power assessment model (P4) where both series come from non-linear specifications.\\

\noindent \textbf{\underline{Size assessment models}}
\begin{description}
  \item[S1]  $X_t=0.5X_{t-1}+\eta_t$, $Y_t=0.5Y_{t-1}+ \epsilon_t$
  \item[S2]  $X_t=\lvert X_{t-1}\rvert^{0.8}+\eta_t$, $Y_t=0.5Y_{t-1}+ \epsilon_t$
  \item[S3]  $X_t=0.5X_{t-1}+\eta_t$, $Y_t=0.5Y_{t-1}+0.5X_{t-1}^2+ \epsilon_t$
  \item[S4] $X_t=0.5X_{t-1}\exp\{-0.5X_{t-1}^2\}+\eta_t$, $Y_t=0.5Y_{t-1}+ \epsilon_t$
  \item[S5] $X_t=\sin{(X_{t-1})}+\eta_t$, $Y_t=0.5Y_{t-1}+ \epsilon_t$
\end{description}

\noindent \textbf{\underline{Power assessment models}}

\begin{description}
  \item[P1] $X_t=0.5X_{t-1}+0.5Y_{t-1}+\eta_t$, $Y_t=0.5Y_{t-1}+ \epsilon_t$ 
  \item[P2] $X_t=0.5X_{t-1}+0.5Y_{t-1}+0.5\sin{(-2Y_{t-1})}+\eta_t$, $Y_t=0.5Y_{t-1}+ \epsilon_t$ 
  \item[P3] $X_t=0.5X_{t-1}+0.5Y_{t-1}^2+\eta_t$, $Y_t=0.5Y_{t-1}+\epsilon_t$  
  \item[P4] $X_t=0.5X_{t-1}+0.5Y_{t-1}^4+\eta_t$, $Y_t=0.5\sin{(Y_{t-1})}+\epsilon_t$ 
  \item[P5]  $X_t=0.65X_{t-1}+0.2Y_{t-1}^2+\eta_t$, $Y_t=-0.3Y_{t-1}+\epsilon_t$
\end{description}

\noindent where $(\epsilon_t,\eta_t)$ are i.i.d. from a standard bivariate normal distribution. 
Recall the size of a test corresponds to the probability of incorrectly rejecting the null hypothesis when it is true, 
whilst the power of a test is the probability of correctly rejecting the null hypothesis when it is false. 
Consequently, the size assessment models correspond to data-generating processes for which there is no Granger causality 
from $Y$ to $X$ (note that S3 exhibits Granger causality from $X$ to $Y$, but not from $Y$ to $X$), 
whereas the models used for the power assessment of the test are built under the presence of Granger causality 
from $Y$ to $X$.\\

\indent For the simulation study, we used $S=500$ simulations for each model with sample sizes of $T=50$, $T=100$ and $T=200$,  
and computed the empirical size and power using a significance level $\alpha=0.05$. Furthermore, 
for part A of the procedure, we used $N=200$ i.i.d. predictions for each series to estimate the 
conditional expectations and, for part B, i.e. the computation of the $p$-value, we used a number 
$B=200$ of generated samples. We used $T_0=T/2$ in the computation of the test statistic 
in accordance with the choice $T^*=T/2$ in \cite{jang2022}.\\

\begin{table}[H]
\centering
\caption{Empirical size comparison between Granger causality tests 
applied to Markov bivariate stationary process of order $k=1$.}
\resizebox{0.98\textwidth}{!}{
\label{tablesize}
\begin{tblr}{
  row{1} = {c},
  column{4} = {c},
  column{8} = {c},
  column{12} = {c},
  cell{1}{1} = {r=2}{},
  cell{1}{2} = {c=4}{},
  cell{1}{6} = {c=4}{},
  cell{1}{10} = {c=4}{},
  cell{3}{2} = {c},
  cell{3}{3} = {c},
  cell{3}{5} = {c},
  cell{3}{6} = {c},
  cell{3}{7} = {c},
  cell{3}{9} = {c},
  cell{3}{10} = {c},
  cell{3}{11} = {c},
  cell{3}{13} = {c},
  cell{4}{2} = {c},
  cell{4}{3} = {c},
  cell{4}{5} = {c},
  cell{4}{6} = {c},
  cell{4}{7} = {c},
  cell{4}{9} = {c},
  cell{4}{10} = {c},
  cell{4}{11} = {c},
  cell{4}{13} = {c},
  cell{5}{2} = {c},
  cell{5}{3} = {c},
  cell{5}{5} = {c},
  cell{5}{6} = {c},
  cell{5}{7} = {c},
  cell{5}{9} = {c},
  cell{5}{10} = {c},
  cell{5}{11} = {c},
  cell{5}{13} = {c},
  cell{6}{2} = {c},
  cell{6}{3} = {c},
  cell{6}{5} = {c},
  cell{6}{6} = {c},
  cell{6}{7} = {c},
  cell{6}{9} = {c},
  cell{6}{10} = {c},
  cell{6}{11} = {c},
  cell{6}{13} = {c},
  cell{7}{2} = {c},
  cell{7}{3} = {c},
  cell{7}{5} = {c},
  cell{7}{6} = {c},
  cell{7}{7} = {c},
  cell{7}{9} = {c},
  cell{7}{10} = {c},
  cell{7}{11} = {c},
  cell{7}{13} = {c},
  vline{6,10} = {2-7}{},
  hline{1,8} = {-}{},
  hline{2-3} = {2-13}{},
}
Model & T=50   &             &        &        & T=100  &             &        &        & T=200  &             &        &        \\
      & M-vine & Jang et al. & NlinTS & Linear & M-vine & Jang et al. & NlinTS & Linear & M-vine & Jang et al. & NlinTS & Linear \\
S1    & 0.046  & 0.066       & 0.000  & 0.058  & 0.050  & 0.070       &  0.016 & 0.048  & 0.056  & 0.050       &  0.106 & 0.058  \\
S2    & 0.044  & 0.052       & 0.000  & 0.058  & 0.052  & 0.080       & 0.026  & 0.054  & 0.062  & 0.066       &  0.074 & 0.056  \\
S3    & 0.058  & 0.070       & 0.000  & 0.060  & 0.054  & 0.054       &  0.014 & 0.052  & 0.050  & 0.070       &  0.090 & 0.046  \\
S4    &    0.060    &      0.082       &   0.002        &    0.056    &     0.060   &    0.044         &     0.026   &    0.064    &    0.054    &    0.048    & 0.080  & 0.056  \\
S5    &    0.056    &      0.074       &    0.000    &   0.040     &    0.054    &   0.060          &    0.026    &    0.066    &     0.050   &       0.054      &   0.072     &     0.036   
\end{tblr}
}
\end{table}

\begin{table}[H]
\centering
\caption{Empirical power comparison between Granger causality tests 
applied to the Markov bivariate stationary process of order $k=1$.}
\resizebox{0.98\textwidth}{!}{
\label{tablepower}
\begin{tblr}{
  row{1} = {c},
  column{4} = {c},
  column{8} = {c},
  column{12} = {c},
  cell{1}{1} = {r=2}{},
  cell{1}{2} = {c=4}{},
  cell{1}{6} = {c=4}{},
  cell{1}{10} = {c=4}{},
  cell{3}{2} = {c},
  cell{3}{3} = {c},
  cell{3}{5} = {c},
  cell{3}{6} = {c},
  cell{3}{7} = {c},
  cell{3}{9} = {c},
  cell{3}{10} = {c},
  cell{3}{11} = {c},
  cell{3}{13} = {c},
  cell{4}{2} = {c},
  cell{4}{3} = {c},
  cell{4}{5} = {c},
  cell{4}{6} = {c},
  cell{4}{7} = {c},
  cell{4}{9} = {c},
  cell{4}{10} = {c},
  cell{4}{11} = {c},
  cell{4}{13} = {c},
  cell{5}{2} = {c},
  cell{5}{3} = {c},
  cell{5}{5} = {c},
  cell{5}{6} = {c},
  cell{5}{7} = {c},
  cell{5}{9} = {c},
  cell{5}{10} = {c},
  cell{5}{11} = {c},
  cell{5}{13} = {c},
  cell{6}{2} = {c},
  cell{6}{3} = {c},
  cell{6}{5} = {c},
  cell{6}{6} = {c},
  cell{6}{7} = {c},
  cell{6}{9} = {c},
  cell{6}{10} = {c},
  cell{6}{11} = {c},
  cell{6}{13} = {c},
  cell{7}{2} = {c},
  cell{7}{3} = {c},
  cell{7}{5} = {c},
  cell{7}{6} = {c},
  cell{7}{7} = {c},
  cell{7}{9} = {c},
  cell{7}{10} = {c},
  cell{7}{11} = {c},
  cell{7}{13} = {c},
  vline{3,7,11} = {1}{},
  vline{6,10} = {2-7}{},
  hline{1,8} = {-}{},
  hline{2-3} = {2-13}{},
}
Model & T=50   &             &        &        & T=100  &             &        &        & T=200  &             &        &        \\
      & M-vine & Jang et al. & NlinTS & Linear & M-vine & Jang et al. & NlinTS & Linear & M-vine & Jang et al. & NlinTS & Linear \\
P1    & 0.706  & 0.630       & 0.000  & 0.958  & 0.916  & 0.856       & 0.038  & 1.000  & 0.996  & 0.978       &  0.172 & 1.000  \\
P2    & 0.544  & 0.442       &  0.002 & 0.802  & 0.720  & 0.676       &  0.090 & 0.982  & 0.942  & 0.886       & 0.378 & 1.000  \\
P3    & 0.430  & 0.238       & 0.000  & 0.348  & 0.716  & 0.384       & 0.034  & 0.388  & 0.986  & 0.644       &  0.172 & 0.392  \\
P4    & 0.684  & 0.426       &  0.000 & 0.530  & 0.972  & 0.628       &  0.176 & 0.518  & 1.000  & 0.950       & 0.612  & 0.508  \\
P5    &     0.142   &     0.084        &    0.002    &    0.092    &    0.198    &         0.146    &    0.024   &   0.064     &   0.314     &     0.214        &     0.128   &  0.052      
\end{tblr}
}
\end{table}

\indent Table~\ref{tablesize} shows evidence that the M-vine test and the linear test control
the empirical size below or around the significance level of $\alpha=0.05$ better than the S-vine test. Furthermore, the NlinTS test is the one that exhibits higher size distortions; in fact, when compared to the other tests, only the S-vine test by \cite{jang2022} exhibits in some cases a size greater than or equal to $0.070$. In terms of power, the NlinTS test is the one with the worst performance. Moreover,  
  the M-vine test outperforms the S-vine test, in every power assessment  model, in small and large samples, with the biggest difference when working with series from the data-generating model P3, that exhibit a quadratic dependence relationship. Furthermore, the power of the M-vine test seems to increase as the sample size grows faster than the one of the S-vine test. Notice that the M-vine and the S-vine tests have the same value of the test statistic under the null hypothesis, i.e.~$\widehat{GC}^{0\;mean}_{Y\rightarrow X}$, and so the difference in the performance relies upon the computation of $\widehat{GC}^{mean}_{Y\rightarrow X}$, that explains the difference in the obtained $p$-values.  When compared to the linear test, the M-vine test tends to have a smaller power across the assessment models P1 and P2 that correspond, respectively, to  a linear model and to  a non-linear model with a bounded non-linear term; this difference is decreased as we move to larger samples, becoming negligible at $T=200$. On the contrary, the power of M-vine test is substantially bigger than the one of the linear test when working with non-linear  model specifications (precisely, with models exhibiting a quadratic dependence), such as P3, P4, and P5, particularly when working with larger samples: indeed, the power of the linear test does not increase as the sample size grows, remaining considerably low. Notably, all the tests considered in the analysis exhibit low power working with the P5 model, in which the weight of the term $Y_{t-1}$ in the dynamics of the process $X$ is lower than the one in the previous models, which implies that $Y$ Granger-causes $X$ to a lesser extent, requiring a larger sample size to achieve a higher power. Hence,  the presented analyses show that the M-vine test brings remarkable improvements   in the current literature regarding tools for detecting the Granger causality in the presence of non-linear dependencies.
\\

\begin{table}[H]
\centering
\caption{Average $p$-value and their empirical standard deviation (in parentheses) across simulations for the size assessment models ($k=1$).}
\resizebox{0.99\textwidth}{!}{
\label{tableavgpvaluessize}
\begin{tblr}{
  row{odd} = {c},
  row{4} = {c},
  row{6} = {c},
  cell{1}{1} = {r=2}{},
  cell{1}{2} = {c=4}{},
  cell{1}{6} = {c=4}{},
  cell{1}{10} = {c=4}{},
  cell{2}{10} = {c},
  cell{2}{11} = {c},
  cell{2}{13} = {c},
  vline{6,10} = {2-7}{},
  hline{1,8} = {-}{},
  hline{2-3} = {2-13}{},
}
Model & T=50              &                   &                   &                   & T=100             &                    &                   &                    & T=200              &                    &                   &                    \\
      & M-vine            & Jang et al.       & NlinTS            & Linear            & M-vine            & Jang et al.        & NlinTS            & Linear             & M-vine             & Jang et al.        & NlinTS            & Linear             \\
S1    & {0.456\\ (0.249)} & {0.520\\ (0.313)} & {
0.978\\(0.097)} & {0.489\\(0292)}   & {0.462\\ (0.264)} & {0.504\\ (0.305)}  & {
0.912\\(0.222)} & {0.509\\(0.294)}   & {0.476\\(0.276)}   & {0.502\\(0.287)}   & {
0.755\\(0.369)} & {0.494\\(0.285)}   \\
S2    & {0.463\\ (0.260)} & {0.519\\ (0.313)} & {
0.970\\(0.104)} & {0.480\\(0.291)}  & {0.471\\ (0.268)} & {0.501\\ (0.305)}  & {
0.881\\(0.259)} & {0.496\\(0.290)}   & {0.473\\(0.277)}   & {0.538\\(0.311)}   & {
0.798\\(0.344)} & {0.487\\(0.292)}   \\
S3    & {0.474\\ (0.251)} & {0.527\\ (0.304)} & {
0.977\\(0.085)} & {0.495\\(0.293)}  & {0.457\\ (0.267)} & {0.516\\ (0.307)}  & {
0.916\\(0.220)} & {0.483\\(0.282)}   & {0.474\\(0.277)}   & {0.507\\(0.293)}   & {
0.768\\(0.365)} & {0.507\\(0.292)}   \\
S4    & {0.475\\ (0.270)} & {0.481\\(0.296)}  & {0.982\\(0.082)}  & {
0.504\\(0.291)} & {
0.484\\(0.283)} & {
0.495\\(0.293)}  & {0.891\\(0.251)}  & {0.491\\(0.291)}   & {
0.489\\(0.277)}  & {
0.504\\(0.287)}  & {
0.778\\(0.356)} & {
0.490\\(0.296)}  \\
S5    & {0.460\\ (0.259)} & {
0.503\\(0.299)} & {
0.972\\(0.113)} & {
0.484\\(0.289)} & {
0.460\\(0.265)} & {
0.513\\(0.301) } & {
0.874\\(0.266)} & {
0.482\\(0.286) } & {
0.494\\(0.278) } & {
0.504\\(0.301) } & {
0.788\\(0.343)} & {
0.521\\(0.289) } 
\end{tblr}
}
\end{table}

\begin{table}[H]
\centering
\caption{Average $p$-value and their empirical standard deviation (in parentheses) across simulations for the power assessment models ($k=1$).}
\resizebox{0.98\textwidth}{!}{
\label{tableavgpvaluespower}
\begin{tblr}{
  row{odd} = {c},
  row{4} = {c},
  row{6} = {c},
  cell{1}{1} = {r=2}{},
  cell{1}{2} = {c=4}{},
  cell{1}{6} = {c=4}{},
  cell{1}{10} = {c=4}{},
  cell{2}{4} = {c},
  cell{2}{8} = {c},
  cell{2}{12} = {c},
  vline{6,10} = {2-7}{},
  hline{1,8} = {-}{},
  hline{2-3} = {2-13}{},
}
Model & T=50               &                   &                   &                    & T=100               &                    &                   &                     & T=200                &                     &                   &                     \\
      & M-vine             & Jang et al.       & NlinTS            & Linear             & M-vine              & Jang et al.        & NlinTS            & Linear              & M-vine               & Jang et al.         & NlinTS            & Linear              \\
P1    & {0.092*\\ (0.187)} & {0.190\\ (0.308)} & {
0.956\\(0.137)} & {0.011**\\(0.058)} & {0.021**\\ (0.085)} & {0.074*\\ (0.214)} & {0.851\\(0.281)}  & {0.000***\\(0.001)} & {0.001***\\ (0.007)} & {0.014**\\ (0.104)} & {
0.593\\(412)}   & {0.000***\\(0.000)} \\
P2    & {0.159\\ (0.234)}  & {0.290\\ (0.335)} & {
0.954\\(0.151)} & {0.046**\\(0.113)} & {0.093*\\ (0.199)}  & {0.165\\ (0.298)}  & {
0.772\\(0.367)} & {0.004***\\(0.025)} & {0.017**\\ (0.085)}  & {0.050*\\ (0.162)}  & {
0.499\\(0.461)} & {0.000***\\(0.000)} \\
P3    & {0.214\\ (0.268)}  & {0.454\\ (0.356)} & {
0.970\\(0.109)} & {0.294\\(0.309)}   & {0.091*\\ (0.201)}  & {0.348\\ (0.354)}  & {
0.855\\(0.288)} & {0.262\\(0.303)}    & {0.011***\\ (0.089)} & {0.216\\ (0.328)}   & {
0.619\\(0.412)} & {0.245\\(0.277)}    \\
P4    & {0.121\\(0.233)}   & {0.311\\(0.355)}  & {
0.915\\(0.195)} & {0.195\\(0.276)}   & {0.015**\\(0.104)}  & {0.190\\(0.310)}   & {
0.658\\(0.418)} & {0.195\\(0.275)}    & {0.000***\\(0.000)}  & {0.033**\\(0.148)}  & {
0.305\\(0.432)} & {0.266\\(0.305)}    \\
P5    & {0.395\\(0.279)}   & {
0.469\\(0.298)} & {0.972\\(0.109)}  & {0.465\\(0.295)}   & {
0.346\\(0.289)}   & {
0.445\\(0.311)}  & {
0.878\\(0.266)} & {
0.491\\(0.310)}   & {
0.285\\(0.292)}    & {
0.404\\(0.319)}   & {
0.718\\(0.389)} & {0.469\\(0.293)}    
\end{tblr}
}
\begin{tablenotes}
      \small
      \item * $\overline{p} < 0.1$, ** $\overline{p} < 0.05$, *** $\overline{p} < 0.01$, with $\overline{p}$ denoting the average $p$-value.
    \end{tablenotes}
\end{table}

\begin{figure}[H]
\centering
\begin{subfigure}{.5\textwidth}
  \centering
  \includegraphics[width=1\linewidth]{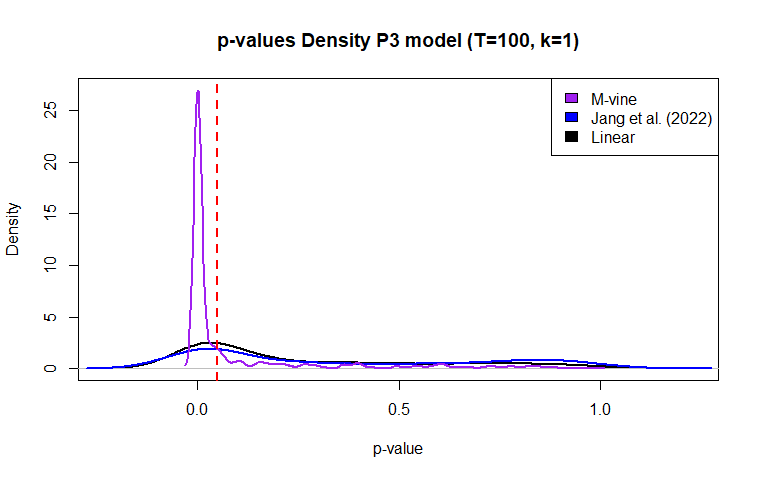}
  
\end{subfigure}%
\begin{subfigure}{.5\textwidth}
  \centering
  \includegraphics[width=1\linewidth]{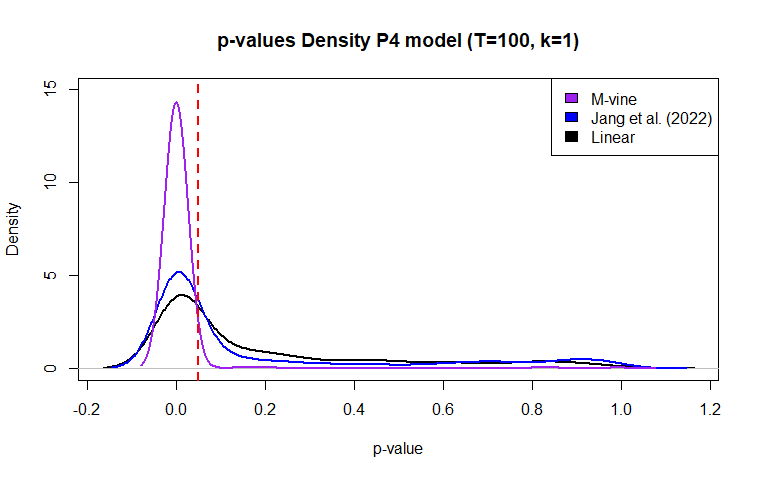}

\end{subfigure}
\caption{Distribution of the $p$-values obtained by each method for both non-linear assessment models P3 and P4, in the case $k=1$ and $T=100$. The red dashed line represents the significance level $\alpha=0.05$.}
\label{Exdens}
\end{figure}

\indent Table~\ref{tableavgpvaluessize} and~\ref{tableavgpvaluespower} show the average $p$-value across simulations for each size and power assessment model, respectively, together with their empirical standard deviation. One can see that, on average, the $p$-values for all methods are around the significance level $\alpha=0.05$ in every size assessment model (with the exception of P5 due to the sample size as previously mentioned). However, those of the M-vine approach have the lower standard deviation among the three tests in all size models, both in small and larger samples. On the other hand, from Table~\ref{tableavgpvaluespower}, we can see that for the linear (P1) and close to linear (P2) model specifications, the $p$-values of the Linear method are the ones that are, on average, further below the predefined significance level, also displaying the lower standard deviation. When working with non-linear dependence structures (P3, P4 and P5), this no longer holds as the $p$-values for the M-vine procedure are, on average, closer to the significance level, and they exhibit the lower empirical deviation out of the three methods. In Figure~\ref{Exdens} we plot the distributions of the $p$-values obtained with each method in cases P3 and P4 with $T=100$ in order to further support the previous considerations.\\

\indent Thus far, we have assumed that $(X_t,Y_t)_t$ is a first-order Markov process (i.e. $k=1$), and, under this condition, the M-vine test results to have good statistical properties for sample size $T\geq 100$. In order to analyse how these properties change when working with higher-order Markov processes, we repeat the simulation study for the case $k=4$. The selection of this particular Markov order is driven by the empirical application from Section~\ref{empapp}, as it is the only other Markov order, besides $k=1$, obtained by our selection procedure when fitting the M-vine models to the U.S. macroeconomic data (we refer to Section~\ref{datadesc} for a detailed description of the data). To adjust the simulation study for $k=4$, we modified a selection of the previously presented size and power assessment models so that they represent the same dependence structure scenarios, but letting $(X_t,Y_t)_t$ be a bivariate Markov stationary stochastic process of order $k=4$. The modified assessment models are given in Appendix~\ref{k4mods}. Table \ref{tablesizek4} and Table \ref{tablepowerk4} show the empirical size and power of each test when applied to a bivariate Markov stationary process of order $k=4$. From them, we can see that overall the results exhibit the same pattern as the case of $k=1$, with the M-vine test still outperforming in terms of power the S-vine test in both linear and non-linear assessment models. Moreover, when compared to the linear test, the M-vine test has a comparable performance when the processes have a linear or almost linear dependence and the sample size is large enough, while it performs better in non-linear cases. However, we note that for the order $k=4$, the power of the M-vine test seems to converge to 1 at a slower pace than in the case of $k=1$. Thus, for $k=4$, we require a greater $T$ to let the M-vine test have a good power in the cases P3 and P4.\\

\begin{table}[H]
\centering
\caption{Empirical size comparison between Granger causality tests applied to a Markov bivariate stationary process of order $k=4$.}
\resizebox{0.9\textwidth}{!}{
\label{tablesizek4}

\begin{tblr}{
  row{1} = {c},
  cell{1}{1} = {r=2}{},
  cell{1}{2} = {c=3}{},
  cell{1}{5} = {c=3}{},
  cell{1}{8} = {c=3}{},
  cell{3}{2} = {c},
  cell{3}{3} = {c},
  cell{3}{4} = {c},
  cell{3}{5} = {c},
  cell{3}{6} = {c},
  cell{3}{7} = {c},
  cell{3}{8} = {c},
  cell{3}{9} = {c},
  cell{3}{10} = {c},
  cell{4}{2} = {c},
  cell{4}{3} = {c},
  cell{4}{4} = {c},
  cell{4}{5} = {c},
  cell{4}{6} = {c},
  cell{4}{7} = {c},
  cell{4}{8} = {c},
  cell{4}{9} = {c},
  cell{4}{10} = {c},
  cell{5}{2} = {c},
  cell{5}{3} = {c},
  cell{5}{4} = {c},
  cell{5}{5} = {c},
  cell{5}{6} = {c},
  cell{5}{7} = {c},
  cell{5}{8} = {c},
  cell{5}{9} = {c},
  cell{5}{10} = {c},
  vline{5,8} = {2-5}{},
  hline{1,3,6} = {-}{},
  hline{2} = {2-10}{},
}
Model & T=50   &                                  &        & T=100  &                                  &        & T=200  &                                  &        \\
      & M-vine & Jang et al. & Linear & M-vine & Jang et al. & Linear & M-vine & Jang et al. & Linear \\
S1    & 0.066  &     0.070                        &  0.040 & 0.048  &         0.040                    & 0.032  & 0.042  &        0.070                     &  0.048 \\
S2    &  0.054 &           0.046                  &  0.050 &   0.048&               0.044              & 0.040  & 0.058  &              0.038               &  0.050 \\
S3    &  0.058 &           0.058                  & 0.052  &  0.054 &           0.088                  & 0.052  &   0.058    & 0.080                                 &      0.060  
\end{tblr}
}
\end{table}

\begin{table}[H]
\centering
\caption{Empirical power comparison between Granger causality tests applied to a Markov bivariate stationary process of order $k=4$.}
\resizebox{0.9\textwidth}{!}{
\label{tablepowerk4}
\begin{tblr}{
  row{1} = {c},
  cell{1}{1} = {r=2}{},
  cell{1}{2} = {c=3}{},
  cell{1}{5} = {c=3}{},
  cell{1}{8} = {c=3}{},
  cell{3}{2} = {c},
  cell{3}{3} = {c},
  cell{3}{4} = {c},
  cell{3}{5} = {c},
  cell{3}{6} = {c},
  cell{3}{7} = {c},
  cell{3}{8} = {c},
  cell{3}{9} = {c},
  cell{3}{10} = {c},
  cell{4}{2} = {c},
  cell{4}{3} = {c},
  cell{4}{4} = {c},
  cell{4}{5} = {c},
  cell{4}{6} = {c},
  cell{4}{7} = {c},
  cell{4}{8} = {c},
  cell{4}{9} = {c},
  cell{4}{10} = {c},
  cell{5}{2} = {c},
  cell{5}{3} = {c},
  cell{5}{4} = {c},
  cell{5}{5} = {c},
  cell{5}{6} = {c},
  cell{5}{7} = {c},
  cell{5}{8} = {c},
  cell{5}{9} = {c},
  cell{5}{10} = {c},
  cell{6}{2} = {c},
  cell{6}{3} = {c},
  cell{6}{4} = {c},
  cell{6}{5} = {c},
  cell{6}{6} = {c},
  cell{6}{7} = {c},
  cell{6}{8} = {c},
  cell{6}{9} = {c},
  cell{6}{10} = {c},
  vline{3,6,9} = {1}{},
  vline{5,8} = {2-6}{},
  hline{1,3,7} = {-}{},
  hline{2} = {2-10}{},
}
Model & T=50   &       &        & T=100  &       &        & T=200  &       &        \\
      & M-vine & Jang et al.  & Linear & M-vine & Jang et al.      & Linear & M-vine & Jang et al.      & Linear \\
P1    &  0.802 & 0.428 &  1.000 &  0.986 & 0.788 &  1.000 & 1.000  & 0.986 & 1.000  \\
P2    & 0.618  & 0.316 &  0.960 &  0.870 & 0.626 & 1.000  & 0.998 & 0.922  & 1.000  \\
P3    & 0.204 & 0.126 & 0.194  &  0.467 & 0.305 &  0.217 & 0.820  & 0.524 & 0.242  \\
P4    &  0.210 & 0.132 & 0.272  & 0.547  & 0.237 & 0.290  & 0.872  & 0.436 & 0.368 
\end{tblr}
}
\end{table}

\section{Application: energy consumption, economic growth and investment} \label{empapp}

The nexus between energy consumption and economic growth is important in times when economic and environmental sustainability are at stake. On the one hand, the correlation is positive everywhere in the world. No wealthy country consumes only a little energy, and no poor country consumes a lot of energy. Yet, beyond correlations, we believe it is important to investigate whether there is any Granger causality between the two variables, especially in a period in which we do have concerns about the impact on the environment and the depletion of natural resources that are needed to generate energy. In the latest decades, since \cite{kraft1978}, the nexus between energy consumption and economic growth has been analysed in a wide array of studies. Nonetheless, results appear to be diverse and often conflicting as some of them find neutrality, whilst others find a causal relationship, often running in different directions. As pointed out in \cite{chen2007}, the main reason for conflicting results might be the perspective taken from different countries that have their own energy policies, institutions, and cultural factors. However, we can find conflicting conclusions even when they refer to the same country. Indeed, the employment of data sets with different proxy variables for energy consumption or different temporal frequencies, as well as different econometric methods, can certainly contribute to different findings about the energy-growth relationship. Therefore, by now, previous literature  \citep{payne2009, ozturk2010} has been inconclusive about policy recommendations for either a specific country or across different nations. Most of these studies apply a notion of Granger causality tests based on linear models, such as Vector Auto Regressive models (VAR). Only a small number of studies experiment with non-linear models (\cite{huang2008}, \cite{chiou2008}, \cite{fallahi2011}). In Appendix~\ref{prevlit}, Table~\ref{table-literature}, we provide a snapshot of the results from different studies that specifically investigated Granger causality between energy consumption and economic growth with a focus on the United States. Finally, we include in our study also the U.S. time series for investment, proxied by using the Gross Fixed Capital Formation, to show how Granger causality can work differently when considering a variable that is not supposed to have a short-term impact on GDP. Crucially, we decide to investigate the U.S. case for two reasons. On the one hand, it is the most investigated case, and we will not find it difficult to compare our results with previous applications. On the other hand, it is easier to find longer time series that are available for our variables of interest.

\subsection{Data}\label{datadesc}

We assemble our data set from different sources. We obtain real GDP  and Gross Fixed Capital Formation (\textit{investment} hereafter) from the Federal Reserve Economic Data (FRED) as made available by the Federal Reserve Bank of St. Louis \footnote{Accessible through https://fred.stlouisfed.org/}. Energy consumption data are proxied by the so-called total primary energy consumption measured in quadrillion British thermal units (BTU), which is made freely available by the U.S. Energy Information Administration (EIA) \footnote{Accessible through https://www.eia.gov/opendata/}. Crucially, we consider all the time series with a quarterly frequency in the period 1973-2018.\\

\begin{table}[H]
\centering
\caption{Variables included in the empirical application and their respective notation.}
\begin{tabular}{@{}lc@{}}
\toprule
Variable                   & Notation                                               \\ \midrule
Real GDP                   &  $Y$ \\
Energy Consumption &    $EC$        \\
Investment            &          $I$             \\ \bottomrule
\end{tabular}
\end{table}

\subsection{Results}

A preliminary investigation of the time series for this empirical application is available in Appendix~\ref{prets}. To proceed, we need stationarity of the time series. Thus, we work with the first differences of original variables after testing that they are indeed stationary by means of the Phillips-Perron test \citep{pp1988}. First of all, we focus on how the fit to the entire data set of an M-vine model compares to both an S-vine model (on which  the test from \cite{jang2022} is based) and a traditional linear model, i.e., a VAR. Table~\ref{table2} shows the results of the fit of these models in terms of the \textit{Akaike Information Criterion} (AIC) for each pair of variables considered in the analysis. The Markov order of each bivariate process is selected by iteration, fitting M-vine copulas with Markov orders from $1$ to $4$, thus keeping the one with the better fit according to the AIC. Analogously, the same process is repeated for the case of S-vine models. For the linear model, we select the optimal amount of lags in the bivariate VAR models based on the AIC. That is, we sequentially fit VAR models with an increasing number of lags, and then we pick the one with the minimum AIC \footnote{Please note that we implement the command \textit{VARselect} from the RStudio package \textit{VAR Modelling (vars)}.}.  \\

\begin{table}[H]
\centering
\caption{Goodness of fit based on AIC for M-vine, S-vine and VAR model specification. The selected Markov order is in parentheses for the vine models.}
\label{table2}
\begin{tblr}{
  cells = {c},
  hline{1-2,5} = {-}{},
}
Variables & M-vine    &   S-vine   & VAR     \\
$(\Delta Y,\Delta EC)$    & {-50.69\\$(k=1)$} & {-50.69\\$(k=1)$}& 2185.22 \\
\hline
$(\Delta Y,\Delta I)$     & {-226.27\\$(k=1)$} & {-226.27\\$(k=1)$} & 3529.93 \\
\hline
$(\Delta EC,\Delta I)$    & {-333.70\\$(k=4)$} &{-273.62\\$(k=4)$} & 1967.58 \\

\end{tblr}
\end{table}

Table~\ref{table2} put in evidence the non-linear dependence structure between every pair of  considered variables: indeed, once we allow for non-linear models, the goodness of fit is considerably higher. Furthermore, when comparing the goodness of fit between the two vine copula models, one can notice that the M-vine and the S-vine models exhibit almost the same AIC. In particular, in the first two cases, the model selected is exactly the same for both vine copula models, suggesting that the M-vine is the structure that has the best goodness of fit among all the vine structures that can represent stationary time series (i.e. S-vine copulas).\\

\begin{table}[H]
\centering
\caption{$p$-values for the pairwise Granger-causality test. We recall that $T=181$.}
\label{table3}
\begin{threeparttable}
\begin{tabular}[t]{llll}
\toprule
Causality flow\\
(from$\rightarrow$to) & $p$ (M-vine) & $p$ (Jang et al.) & $p$ (Linear)\\
\midrule
$\Delta Y\rightarrow \Delta EC$ &0.015**& 0.030** &0.078*\\
\midrule
$\Delta EC\rightarrow \Delta Y$&0.080*& 0.395 &0.206\\
\midrule
$\Delta Y\rightarrow \Delta I$&0.535& 0.710&0.465\\
\midrule
$\Delta I\rightarrow \Delta Y$&0.265& 0.195&0.006***\\
\midrule
$\Delta EC\rightarrow \Delta I$&0.200& 0.140&0.436\\
\midrule
$\Delta I\rightarrow \Delta EC$&0.035**& 0.045**&0.032**\\
\bottomrule

\end{tabular}
\begin{tablenotes}
      \small
      \item * $p < 0.1$, ** $p < 0.05$, *** $p < 0.01$. The null hypothesis of no Granger causality is rejected at a given significance level $\alpha$ if $p<\alpha$.
    \end{tablenotes}
 \end{threeparttable}
\end{table}

\indent The results we obtain with our M-vine test, the test from \cite{jang2022}, and the more traditional test based on a linear model are \textit{prima facie} relatively similar. Please recall that the null hypothesis of these tests is that no Granger causality exists, and it is rejected at a given significance level $\alpha$ if $p<\alpha$. Remarkably, the most important difference in the results is for the pair  GDP and energy consumption. In fact, as Table~\ref{table3} shows, all three methods find that Granger causality runs from GDP to energy consumption, although the linear test shows a relatively weaker statistical significance at $10\%$ level, in line with the findings from~\cite{kraft1978}, \cite{fallahi2011} and \cite{aslan2014}. 

Interestingly, only the M-vine test detects an inverse Granger causality from energy consumption to GDP at the $10\%$ level. In this specific case, findings are in line 
with~\cite{stern1993, stern2000, bowden2009, fallahi2011}. 
As shown in Table~\ref{table2}, the difference we observe when we compare the M-test and the VAR is due to a non-linear dependence structure between the variables. In this case, we argue that the M-vine and the S-vine tests are better than a classical VAR test at catching the underlying functional form. Moreover, as already shown with the simulation study in previous paragraphs, the M-vine test outperforms the S-vine test thanks to the methodological differences we discussed in section~\ref{sec-diff}. Hence, the differences that we record in the $p$-values of Table \ref{table3}. 

\indent Notably, our findings point to a two-way Granger causality between energy consumption and GDP that is important from an economic perspective. Inevitably, energy consumption is a relevant expenditure component for consumers and producers of any budget. For this reason, economic growth is reflected necessarily on increasing energy consumption. From another point of view, we can say that the demand for energy is quite rigid for both consumers and producers.

Please note how another important difference arises when we test the pairwise causal relationship between investment and economic growth. In this case, the linear test rejects the null hypothesis, thus finding evidence of a Granger causality that runs from investment to GDP, whilst both non-linear tests find no Granger-causal relationship in either direction. In this regard, please consider that Table~\ref{table2} shows the VAR model has absolutely the worst fit when modelling the dependence structure between investment and economic growth. Once again, we believe this is a hint that the results of non-linear tests are more reliable when applied to relationships that can hardly be proxied as linear. From an economic perspective, we know that investment might not have a direct effect on GDP, but it can contribute indirectly to economic growth in the medium term after raising the technological abilities of a country and, thus, its production.  

Finally, all three methods find evidence that investment Granger-causes energy consumption growth. In this case, we believe that the tests are catching the intrinsic prediction power of investment plans that always require the use of additional energy resources.\\

\section{Concluding remarks}\label{conc}

Motivated by the recent literature about non-linear Granger causality tests, our aim was to construct a semi-parametric Granger causality-in-the-mean test for bivariate $k$-Markov stationary processes based on vine copulas. Departing from the procedure in \cite{jang2022}, we added coherence between the two parts of the test, i.e, the estimation of the Granger causality measure and the computation of the $p$-value, by fitting the same M-vine structure in both parts. In addition, we modified the approach by fitting the vine copula model to the entire sample. Therefore, by means of a simulation study with time series from different data generating processes and Markov orders, we showed that we improve on the statistical properties, making the proposed M-vine test an excellent tool for testing Granger causality in the presence of non-linear dependence structures.\\

\indent Finally, in an application, we use the M-vine test to study the relationship between energy consumption, economic growth and investment in the U.S. Crucially, the copula-based M-vine test revealed a two-way Granger causality between energy consumption and GDP that was not detected by the other tests. We argue that, indeed, a quite rigid demand for energy can be the basis for the two-way prediction power in the relationship.

\indent More in general, we conclude that vine copulas are able to model the dependence structure of multiple stochastic processes, better catching intrinsic non-linearities. A possible avenue for future research is the extension of vine copula-based tests to multivariate settings.

\section*{Acknowledgments}

Irene Crimaldi is partially supported by the project MOTUS - Automated Analysis and Prediction of Human Movement Qualities (code P2022J8AXY, cup: D53D23017470001), financed by the Italian Ministry of University and Research (PRIN 2022 PNRR)
\appendix
\section{Appendix}
\subsection{Vine copula models}\label{copapp}
Copula functions are the mathematical tool that describes any type of dependence structure. Hence, in contrast to VAR, copula models 
%
%
allow for non-linear dependences when working with high dimensional data. 

Moreover, Sklar's theorem \citep{sklar1959fonctions} shows that the marginal distributions of the random variables in a multivariate system 
can be selected freely and then the dependence structure among these marginals can be modeled 
by linking them through an appropriate copula function.
\begin{theorem}[Sklar's Theorem]
Consider a $d$-dimensional joint cumulative distribution function $H$ with one-dimensional marginals $F_1,\dots,F_d$. Then there exists a d-copula $C$ such that for $(x_1,\dots,x_d)\in \mathbb{R}^d$,\[H(x_1,\dots,x_d)=C(F_1(x_1),\dots,F_d(x_d)).\]
On the other hand, the converse is also true, i.e. consider a copula $C$ and univariate cumulative distribution functions $F_1,\dots,F_d$. Then $H$ is a multivariate cumulative distribution function with marginals $F_1,\dots,F_d$.
\end{theorem}
However, copula functions have their own limitations when we have to select and estimate them by means of the data. 
Particularly, when dealing with high-dimensional random vectors the set of available parametric copulas is considerably limited and so the selection of a suitable high-dimensional copula function could result difficult. {\em Vine copulas} come as an alternative:  since the large amount of existing bivariate copula families to be employed, the vine copula approach consists in the decomposition of a multivariate copula density into a product of 
bivariate copula densities through the use of the pair-copula construction (\cite{joe1996}). 
Nevertheless, the number of possible decompositions becomes difficult to track and organize: in fact, the number of possible decompositions increases super-exponentially with the dimension of the random vector \citep{moralesnapoles2011}.  
Then, \emph{Regular vines} (R-vines) were introduced as a graphical way of specifying the sequence of trees that characterises 
all valid factorizations of a $d$-dimensional copula. 
Briefly, as defined on \cite{kurowickacooke2006} 
an R-vine on $d$ random variables is a set of graphs 
$\mathcal{V}=(T_1;\dots,T_{d-1})$ where each tree $T_i$ has a node set $N_i$ and an edge set $E_i$, such that:
\begin{enumerate}
  \item $T_1$ has node set $N_1={1,\dots,d}$ and edge set $E_1$. Namely, each node of $T_1$ is associated to one variable from the random vector whose dependence is being modeled.
  \item For $i\geq 2$, $T_i$ has node set $N_i=E_{i-1}$.
  \item Proximity condition: if two edges in tree $T_i$ are joined as nodes in the following tree $T_{i+1}$, then they must share a common node in $T_i$.
\end{enumerate}

\noindent Each edge $e\in E_i$ in any of the trees of a given R-vine is associated to a bivariate copula $C_{a_e,b_e|D_e}$ and three subsets of the nodes from the first tree $T_1$, they are:
\begin{itemize}
  \item $D_e$: the conditioning set of $e$.
  \item $\{a_e,b_e\}$: the conditioned set of $e$ which contains necessarily two elements.
  \item $U_e$: the complete union of $e$, that is $U_e=\{a_e,b_e,D_e\}$.
\end{itemize}

\begin{figure}[H]
\centering
\includegraphics[width=0.65\linewidth]{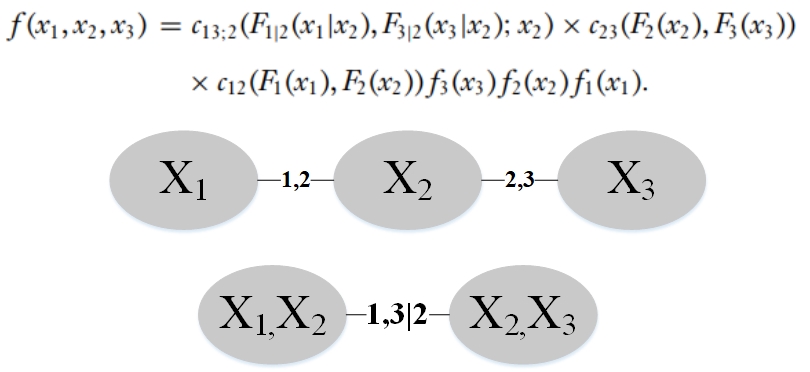}
\caption{Graphical representation of a possible R-vine decomposition in dimension three.}
\label{vine1}
\end{figure}

\noindent Figure~\ref{vine1} provides an example of one possible pair copula decomposition for a three dimensional random vector, 
which is going to be used to clarify the introduced concepts. To begin with, notice that, 
since we are working in dimension three, the total number of trees is 2. The first tree $T_1$ has node set $N_1=\{1,2,3\}$ and 
edge set $E_1=\{\{1,2\},\{2,3\}\}$, whereas the second tree $T_2$ has $N_2=E_1=\{\{1,2\},\{2,3\}\}$ 
and $E_2=\{\{1,3|2\}\}$ respectively. Furthermore, let us consider the edge $\{1,3|2\}\in E_2$ which has conditioning set $D_e=\{2\}$, 
conditioned set $\{1,3\}$ and thus its complete union is $U_e=\{1,3,2\}$.

\noindent To generalise, for a $d$-dimensional random vector $\bm{X}=(X_1,\dots,X_d)$, 
its joint density can be expressed as the following product of$\;\frac{d(d-1)}{2}$ bivariate copula densities and of all the marginal densities:
\begin{equation*}
f(x_1,\dots,x_d)=\prod_{i=1}^{d-1}\prod_{e\in E_i}c_{{a_e,b_e|D_e}}(F_{a_e|D_e},F_{b_e|D_e})\times\prod_{i\in N_1}f_i \; \text{.}
\end{equation*}

\subsubsection{Vine copula models for stationary time series: S-vine copulas}

\noindent Remarkably, R-vines are only able to capture the cross-sectional dependence of a random vector: indeed, they model the joint distribution of a random vector $\bm{X}=(X_1,\dots,X_d)$, leveraging on $T$ independent observations of it. Therefore, afterwards, vine copula specifications for modeling time series that are able to capture both cross-sectional and serial dependence 
have been developed (\cite{brechmannczado2015}, \cite{beare2015}, \cite{smith2015} and \cite{nagler2022}). In particular, the latter work (\cite{nagler2022}) builds upon the previous papers 
to generalize all the vine copula structures that can guarantee strict stationarity of a time series. Instead of 
considering a random vector whose observations are assumed independent over time,  
we take a stationary time series $\bm{X}=(\bm{X}_t)_{t=1,\dots,T}=(X_{1,t},\dots,X_{d,t})_{t=1,\dots,T}$ for which we aim 
to capture its cross-sectional and serial dependence through a vine copula specification. 
Therefore, we now aim to model a joint distribution in the form of 
$F_{1,\dots,d}(x_{1,1},\dots,x_{d,1},\dots,x_{1,T},\dots,x_{d,T})$. 
Note that, before each node from the first tree $T_1$ corresponds  to one of the considered random variables ($X_i$); whereas the nodes from the first tree are now  associated to a tuple $(i,t)$ corresponding to the random variable $X_{i,t}$, where $t$ represents the time period and $i$ is the variable index. The same applies for the elements of the conditioned set $\{a_e,b_e\}$ and the conditioning set $D_e$ 
associated to each edge $e\in E_p$.\\

\indent Graphically, the structure of these models can be seen as a vine that captures the cross sectional dependence of 
each vector $\mathbf{X}_t\in \mathbb{R}^d$ $\forall t=1,\dots,T$.  
Then, each of these cross-sectional vines are connected between 
two consecutive time periods $t$ and $t+1$ through one edge linking one vertex from each of them. {\em S-vine copulas}  allow 
for arbitrary cross-sectional structures and also an arbitrary selection of the variable that is going to connect two 
consecutive cross-sectional vines. In terms of estimation, S-vines select the edge between the two variables that have the highest 
correlation between time periods.
\\

\subsubsection{A sub-class of S-vine copulas: M-vine copulas}
\label{sec-M-vine}

M-vine copula models were one of the first vine copula structures able to represent time series by accounting for both cross sectional and serial dependence. They were first introduce in \cite{beare2015}, particularly, in order to model stationary multivariate higher order Markov chains. Formally, a regular vine $\mathcal{V}$ on the node set $N_1=\{1,\dots,d\}\times\{1,\dots,T\}$ with trees $T_p=(N_p,E_p)$, $p=1,\dots,(dT-1)$, is an \emph{M-vine} if the following conditions are satisfied:
\begin{enumerate}
    \item $E_1=\left\{\{(i,s),(j,t)\}\in \binom{N_1}{p}:(i=j-1 \land s=t)\lor(i=j=s=t-1)\right\}.$
    \item For each pair of adjacent columns $A_t=\cup_{i=1}^{d}\{(i,t),(i,t+1)\}$, $t=1,\dots,n-1$, the \textit{restriction} of $\mathcal{V}$ to $A_t$ is a \emph{Drawable vine} (D-vine).

\end{enumerate}

\noindent where the \textit{restriction}\footnote{For a formal definition we refer to Definition 6 in \cite{beare2015}.} of $\mathcal{V}$ to $A_t$ corresponds to the structure that is left after all nodes that are not included in $A_t$ are deleted from $N_1$, as well as all the edges in the first tree of $\mathcal{V}$ that are connected to these nodes and the edges in subsequent trees that would include them.\\

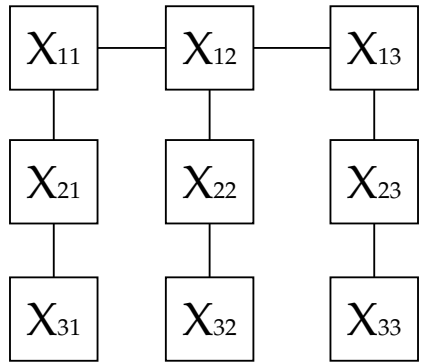
\begin{figure}[H]

\centering    
\resizebox{0.38\textwidth}{!}{

\begin{tikzpicture}[x=0.75pt,y=0.75pt,yscale=-1,xscale=1]

\draw    (27.99,10.01) -- (52.99,10.01) -- (52.99,34.01) -- (27.99,34.01) -- cycle  ;
\draw (40.49,22.01) node   [align=left] {X{\tiny 11}};

\draw    (27.99,48.51) -- (52.99,48.51) -- (52.99,72.51) -- (27.99,72.51) -- cycle  ;
\draw (40.49,60.51) node   [align=left] {X{\tiny 21}};

\draw    (27.99,88.01) -- (52.99,88.01) -- (52.99,112.01) -- (27.99,112.01) -- cycle  ;
\draw (40.49,100.01) node   [align=left] {X{\tiny 31}};
\draw    (72.49,10) -- (97.49,10) -- (97.49,34) -- (72.49,34) -- cycle  ;
\draw (84.99,22) node   [align=left] {X{\tiny 12}};
\draw    (72.49,48.5) -- (97.49,48.5) -- (97.49,72.5) -- (72.49,72.5) -- cycle  ;
\draw (84.99,60.5) node   [align=left] {X{\tiny 22}};
\draw    (72.49,88) -- (97.49,88) -- (97.49,112) -- (72.49,112) -- cycle  ;
\draw (84.99,100) node   [align=left] {X{\tiny 32}};
\draw    (119.49,10) -- (144.49,10) -- (144.49,34) -- (119.49,34) -- cycle  ;
\draw (131.99,22) node   [align=left] {X{\tiny 13}};
\draw    (119.49,48.5) -- (144.49,48.5) -- (144.49,72.5) -- (119.49,72.5) -- cycle  ;
\draw (131.99,60.5) node   [align=left] {X{\tiny 23}};
\draw    (119.49,88) -- (144.49,88) -- (144.49,112) -- (119.49,112) -- cycle  ;
\draw (131.99,100) node   [align=left] {X{\tiny 33}};
\draw    (40.49,34.01) -- (40.49,48.51) ;
\draw    (40.49,72.51) -- (40.49,88.01) ;
\draw    (84.99,34) -- (84.99,48.5) ;
\draw    (84.99,72.5) -- (84.99,88) ;
\draw    (131.99,34) -- (131.99,48.5) ;
\draw    (131.99,72.5) -- (131.99,88) ;
\draw    (52.99,22.01) -- (72.49,22) ;
\draw    (97.49,22) -- (119.49,22) ;

\end{tikzpicture}
}
\caption{First tree of a three-dimensional M-vine on three time periods.}
\label{exmvine}
\end{figure}

\indent Graphically, M-vines can be seen as cross-sectional vines that are connected to the vine associated to the following (and/or previous) time period through an edge linking two nodes that lie in the same end of the vine, i.e between nodes $(1,t)$ and $(1,t + 1)$ as represented in Figure~\ref{exmvine}. These class of structures are able to guarantee stationarity of a multivariate time series, in fact Theorem 3 from \cite{beare2015} shows that a multivariate time series that realises a M-vine copula specification is stationary if and only if the M-vine structure is \textit{translation invariant}. Intuitively, translation invariance requires that the cumulative distribution functions associated to any two nodes in the same row of the first tree structure are the same, and that the same copula is assigned to any two edges whose conditioned and conditioning set differs just by column translations, i.e translations across time. Additionally, M-vines can represent $k$-Markov stationary time series, imposing the Markov property by requiring certain copulas from the vine specification to be independence copulas. For a formal definition of translation invariance and the required conditions to ensure the $k$-Markov property of a given time series using M-vine structures, we refer to Definition 9 and Theorem 4 from \cite{beare2015}, respectively.

\indent The main difference between M-vine and S-vine copulas relies on the fact that the tree structures for a M-vine are fixed as the cross-sectional vines have to be connected across time periods through an edge between two nodes lying in the same end of the vine. On the other hand, S-vines allow for general cross-sectional structures (not necessarily D-vines) and provide a higher degree of freedom for connecting them across time compared to M-vines. This indeed provide more flexibility for S-vines in order to fit different kind of data; however, in cases when we need to guarantee the presence of one or some specific copulas in the vine structure, this can be an undesirable feature: for instance, in Part B of our statistical procedure illustrated in   Section~\ref{mvtest}, where the copulas $c_{X_{t},X_{t+1}}$ and $c_{X_t,Y_t}$ are needed to generate  independent samples under the null hypothesis, using an S-vine structure is not feasible as these two copulas might not appear in its first tree. Therefore, even if S-vines constitute a  general class of vine copulas that can ensure stationarity of a time series, in which M-vines are a sub-class, in some cases when we need to detect particular dependencies, fitting a more restrictive vine structure (e.g.~an M-vine) could be beneficial.

\subsubsection{Selection and estimation for M-vine copulas}
\label{sec-estimation-details}

In order to select the random variables associated to each node of the vine structure, and the pair copula families associated to each edge in the vine structure, one could rely on a preliminary analysis of the data being used. In practice, one could select which variables are associated to each node in the first tree considering the scope of the particular application considered, and then select the copula families associated to each edge using scatter plots of pair of empirical probability integral transforms of the corresponding variables and their lagged values \citep{beare2015}.\\

Alternatively, one could use the algorithm from \cite{dissmann2013} in order to select and estimate a vine copula specification, which first fits the strongest dependence based on the absolute value of the empirical Kendall's $\tau$ between any pair of random variables in the system and, subsequently, the selection of the pair copulas and the estimation of their corresponding parameters is done sequentially through step-wise maximum likelihood estimation (MLE). Firstly, the empirical distribution functions are used to estimate the marginals by

$$
\widehat{F}_j(x_j)= \frac{1}{T+1}\sum_{t=1}^{T}\mathbb{1}(X_{t,j}\leq x),\qquad t=1,\dots,T, j=1,\dots,d.
$$

Using these estimates, one can obtain "pseudo-observations" $\widehat{U}=\widehat{F}_j(X_j)$, that are used for the estimation of the copula parameters. The step-wise MLE estimates the parameters of the pair-copulas of each tree $T_k$ separately, beginning with the first tree. For the setting of a $k$-Markov process, \cite{nagler2022} showed that this can be done through

$$
\widehat{\boldsymbol\theta}_{[e']}=  \arg \max_{\boldsymbol\theta_{\mathbf{[e']}}} \sum_{e\sim e'} \log c_{[e]}\left\{ C_{a_{[e]}|D_{[e]}}\left( \widehat{U}_{a_e} |\mathbf{\widehat{U}}_{D_e}; \widehat{\boldsymbol\theta}_{S_a([e])}, \right),C_{b_{[e]}|D_{[e]}}\left( \widehat{U}_{b_e} |\mathbf{\widehat{U}}_{D_e}; \widehat{\boldsymbol\theta}_{S_b([e])}, \right) ; \widehat{\boldsymbol\theta}_{[e']} \right\},
$$
where $\boldsymbol\theta_{[e]}$ is the parameter of the copula associated to the edge $e$, $[e]=\{e'\sim e\}$ is the equivalence class of edge $e$, containing all the the edges that share the same copula and parameters than $e$. Lastly, $\widehat{\boldsymbol\theta}_{S_a([e])}$ and $\widehat{\boldsymbol\theta}_{S_b([e])}$ contain previously estimated parameters from prior trees.
\\

\cite{nagler2022} studied the asymptotics properties of this Semi-Parametric (SP) estimator, establishing  suitable assumptions under which the semi-parametric estimate $\widehat{\boldsymbol\theta}^{(SP)}$ is a consistent estimator of the true parameter $\widehat{\boldsymbol\theta}^*$. Specifically, Theorem 6 in that paper states that under two technical conditions (for a formal definition of them, we refer to conditions (SP1) and (SP2) in Section 4.3.2 of \cite{nagler2022}), it holds $\widehat{\boldsymbol\theta}^{(SP)}  \overset{P}{\to}\widehat{\boldsymbol\theta}^*$. Intuitively, the first condition guarantees the identification of the vine copula model parameters, whilst the second one is a classic regularity assumption in semi-parametric copula models for dealing with the explosive behavior of derivatives of copula functions in the corners of the unit hypercube. To ensure the suitability of these assumptions, we worked with the R package \textit{svines} by T.~Nagler, which considers parametric copula families without tail dependence, and with tail dependence in either one, two or four tails.

\subsection{Linear bivariate Granger causality test}\label{lintest}

Let $X=(X_t)_t$, $Y=(Y_t)_t$ be two stationary time series with $t=1,\dots,T$. To test whether $Y$ Granger-causes $X$, the linear test assumes a particular autoregressive lag length $p$, that is 

$$
X_t=c_1+ \sum_{j=1}^p\alpha_jX_{t-j}+\sum_{j=1}^p\beta_jY_{t-j}+u_t\,,
$$
  where $u_t$ is a white noise, and use estimation by Ordinary Least Squares (OLS). Then, an asymptotic $F$ test of the null hypothesis
$$
H_0: \beta_1=\dots=\beta_p=0
$$
is conducted. To implement this test, one needs to compute the sum of squared residuals
$$
RSS_1=\sum_{j=1}^T\hat{u}_t
$$
and compare it with the sum of squared residuals of the univariate autoregressive model of order $p$ for $X$,
$$
RSS_0=\sum_{j=1}^T\hat{e}_t
$$
where
$$
X_t=c_0+\sum_{j=1}^p\gamma_jX_{t-j}+e_t
$$
is also estimated by OLS. Finally, if 
$$
S=\frac{T(RSS_0-RSS_1)}{RSS_1}
$$
is greater than the 5\% critical value for a $\mathcal{X}^2(p)$ distribution, then we reject the null hypothesis that Y does not Granger-cause X.
\subsection{Size and power assessment models for $k=4$}\label{k4mods}

\noindent \textbf{\underline{Size assessment models}}
\begin{description}
  \item[S1]  $X_t=0.5\sum\limits_{p=1}^{4}(-1)^{p+1}X_{t-p}+\eta_t$, $Y_t=0.5\sum\limits_{p=1}^{4}(-1)^{p+1}Y_{t-p}+ \epsilon_t$
  \item[S2]  $X_t=\sum\limits_{p=1}^{4}(-1)^{p+1}\lvert X_{t-p}\rvert^{0.8}+\eta_t$, $Y_t=0.5\sum\limits_{p=1}^{4}(-1)^{p+1}Y_{t-p}+ \epsilon_t$
  \item[S3]  $X_t=0.5\sum\limits_{p=1}^{4}(-1)^{p+1}X_{t-p}+\eta_t$, $Y_t=0.5\sum\limits_{p=1}^{4}(-1)^{p+1}Y_{t-p}+0.5\sum\limits_{p=1}^{4}(-1)^{p+1}X^2_{t-p}+ \epsilon_t$
\end{description}

\noindent \textbf{\underline{Power assessment models}}

\begin{description}
  \item[P1] $X_t=0.5\sum\limits_{p=1}^{4}(-1)^{p+1}X_{t-p}+0.5\sum\limits_{p=1}^{4}Y_{t-p}+\eta_t$, $Y_t=0.5\sum\limits_{p=1}^{4}(-1)^{p+1}Y_{t-p}+ \epsilon_t$ 
  \item[P2] $X_t=0.5\sum\limits_{p=1}^{4}(-1)^{p+1}X_{t-p}+0.5\sum\limits_{p=1}^{4}Y_{t-p}+0.5\sum\limits_{p=1}^{4}\sin(-2Y_{t-p})+\eta_t\text{,}$\\
  
\hspace{-3mm}$Y_t=0.5\sum\limits_{p=1}^{4}(-1)^{p+1}Y_{t-p}+ \epsilon_t$

  \item[P3] $X_t=0.5\sum\limits_{p=1}^{4}(-1)^{p+1}X_{t-p}+0.5\sum\limits_{p=1}^{4}Y^2_{t-p}+\eta_t$, $Y_t=0.5\sum\limits_{p=1}^{4}(-1)^{p+1}Y_{t-p}+ \epsilon_t$ 
 
  \item[P4] $X_t=0.5\sum\limits_{p=1}^{4}(-1)^{p+1}X_{t-p}+0.5\sum\limits_{p=1}^{4}Y^4_{t-p}+\eta_t$, $Y_t=0.5\sum\limits_{p=1}^{4}(-1)^{p+1}\sin(Y_{t-p})+ \epsilon_t$
\end{description}

\noindent where $\epsilon_t,\eta_t$ are i.i.d from a standard bivariate normal distribution.

\subsection{Previous literature:  Granger-causality relationship between Energy Consumption and Economic Growth}\label{prevlit}

We here provide a summary of the results provided by the previous literature related to the Granger causality between energy consumption and economic growth in the United States.\\

\begin{table}[H]
\centering
\caption{Energy Consumption ($EC$) - Economic Growth ($\Delta Y$) Granger-causality  relationship found in previous literature using U.S data with the models used for their respective test. }
\label{table-literature}
\begin{tabular}[t]{lccc}
\toprule
Author(s)&Time period&Model&Causality flow\\
\midrule
\cite{kraft1978}&1946-1974&VAR&$\Delta Y\rightarrow EC$\\
\cite{eden1984}&1974-1990&VAR&$\Delta Y\nleftrightarrow EC$\\
\cite{stern1993}&1947-1990&VAR&$EC\rightarrow \Delta Y$\\
\cite{stern2000}&1948-1994&VAR&$EC\rightarrow \Delta Y$\\
\cite{payne20092}&1949-2006&VAR&$\Delta Y\nleftrightarrow EC$\\
\cite{bowden2009}&1949-2006&VAR&$EC\rightarrow \Delta Y$\\
\cite{fallahi2011}&1960-2005&Markov switching VAR&$\Delta Y\leftrightarrow EC$\\
\cite{aslan2014}&1973-2013&Wavelet analysis and VAR&$\Delta Y\rightarrow EC$\\
\bottomrule
\end{tabular}

\begin{minipage}{12cm}

\vspace{0.1cm}

\vspace{0.1cm}

\small  Notes: Given two variables $A$ and $B$, notation $A\rightarrow B$ refers to a unidirectional flow of Granger-causality from variable $A$ to $B$, $A\leftrightarrow B$ illustrates a bidirectional flow of Granger-causality between both variables, and $A\nleftrightarrow B$ refers to the case of no Granger causality between them in any directions.

\end{minipage}

\end{table}

\subsection{Preliminary time series analysis for the application}\label{prets}

\begin{figure}[H]
\centering
\includegraphics[width=0.9\linewidth]{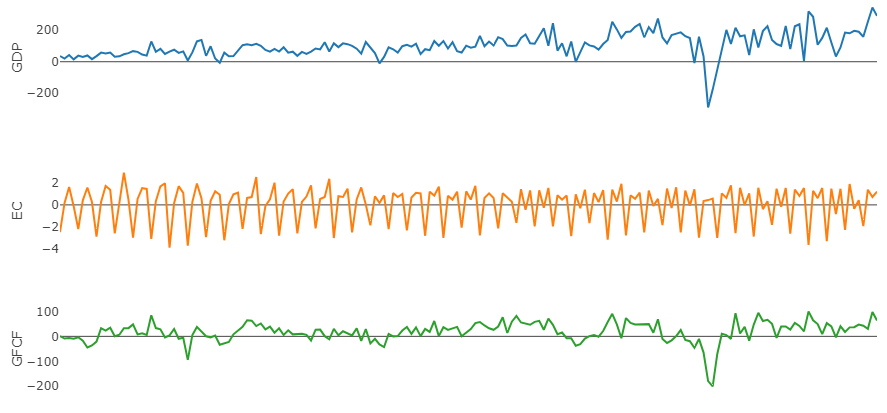}
\caption{Plot of each series in the considered data set in first differences.}
\label{plotts}
\end{figure}

As presented in Figure~\ref{plotts}, one can observe that each variable that is considered in the analysis seems to be stationary, in first differences, around their respective means. \\ 

Moreover, we test the presence of unit roots through a Phillips-Perron test \citep{pp1988} applied to each series in first differences. Recall, that the null hypothesis of the Phillips-Perron test is that the series has a unit root. Given the results in Table~\ref{pptest}, we are able to see that the null hypothesis is rejected at a $\alpha=5\%$ significance level, supporting the previous intuition that every series considered in first differences is stationary. \\

\begin{table}[H]
\centering
\caption{$p$-value for the Phillips-Perron test for each time series in first differences.}
\label{pptest}
\begin{tblr}{
  cells = {c},
  hline{1-2,5} = {-}{},
}
Variable & $p$-value         \\
$\Delta Y$      & $<0.01$ \\
$\Delta EC$       & $<0.01$ \\
$\Delta I$        & $<0.01$            \\
       
\end{tblr}
\end{table}

\bibliography{ms}
\end{document}